\documentclass[aps,letterpaper,showkeys,footinbib,twocolumn]{revtex4-1}
	\usepackage{cancel}
	\usepackage{amsmath,amssymb} 
	\usepackage{amsfonts}
	\usepackage{inputenc}
	\usepackage{subfigure}
	\usepackage{epsfig}
	\usepackage{pst-grad} 
	\usepackage{comment}
	\usepackage[normalem]{ulem}
	\usepackage{natbib}
	\usepackage{footnote}
	
\begin{document}

\title{Scale-invariant perturbations in ekpyrotic cosmologies without fine-tuning of initial conditions}

\author{Aaron M. Levy}
\affiliation{Department of Physics, Princeton University, Princeton, NJ 08544, USA}
\author{Anna Ijjas}
\email{aijjas@princeton.edu}
\affiliation{Princeton Center for Theoretical Science, Princeton University, Princeton, NJ 08544, USA}
\author{Paul J. Steinhardt}
\affiliation{Department of Physics, Princeton University, Princeton, NJ 08544, USA}
\affiliation{Princeton Center for Theoretical Science, Princeton University, Princeton, NJ 08544, USA}

\date{\today}

\begin{abstract}
Ekpyrotic bouncing cosmologies have been proposed  as alternatives to inflation. In these scenarios, the universe is smoothed and flattened during a period of slow contraction preceding the bounce while quantum fluctuations generate nearly scale-invariant super-horizon perturbations that seed structure in the post-bounce universe.  An analysis by Tolley and Wesley (2007) showed that, for a wide range of ekpyrotic models, generating a scale-invariant spectrum of adiabatic or entropic fluctuations is only possible if the cosmological background is unstable, in which case the scenario is highly sensitive to  initial conditions.   In this paper, we analyze an important counterexample: a simple action that  generates a Gaussian, scale-invariant spectrum of entropic perturbations during ekpyrotic contraction without requiring fine-tuned initial conditions.  Based on this example, we discuss some generalizations.
\end{abstract}

\keywords{cyclic/ekpyrotic cosmology, entropic mechanism, attractor solutions}

\maketitle
\section{Introduction}
Observations of the cosmic microwave background from the {\it Wilkinson Microwave Anisotropy Probe} (WMAP) \cite{Komatsu:2008hk}, the {\it Planck} satellite \cite{Ade:2015lrj,Ade:2015ava}, {\it the Atacama Cosmology Telescope} (ACT) \cite{Sievers:2013ica} and other experiments have shown that the primordial scalar (density) fluctuation spectrum is adiabatic and nearly scale-invariant with nearly Gaussian statistics.  Inflation \cite{Guth:1980zm,Albrecht:1982wi,Linde:1981mu} has been suggested as a mechanism for generating  perturbations with these properties, though, it is known that to do so it requires rare initial conditions \cite{Penrose:1988mg,Gibbons:2006pa} and results in a multiverse of outcomes  \cite{Steinhardt:1982kg,Vilenkin:1983xq,Guth:2000ka,Guth:2013sya,Linde:2014nna}.  

Bouncing cosmologies with a period of ultra-slow (ekpyrotic) contraction have been proposed as alternatives.  In these theories, smoothing contraction occurs because the energy density of a scalar field with equation-of-state $\epsilon > 3$ (where $\epsilon \equiv 3(p+\rho)/\rho$  with $p$ being the pressure and $\rho$ the energy density) grows to dominate all other forms of energy, including inhomogeneities, anisotropy and spatial curvature \cite{Khoury:2001wf}.  A key advantage compared to inflation is that the ekpyrotic mechanism does not lead to a multiverse. 

The currently best understood way to produce density fluctuations in the ekpyrotic theory involves two scalar fields that generate a  scale-invariant spectrum of entropy perturbations.   After the ekpyrotic smoothing phase, these perturbations convert into a scale-invariant spectrum of adiabatic perturbations  \cite{Buchbinder:2007ad,Lehners:2008vx,Lehners:2007ac, DiMarco:2002eb}.  In the first examples discussed in the literature, the background cosmological solution describing the evolution of the two fields along the potential energy surface is unstable, which means finely-tuned initial conditions are required to begin the ekpyrotic phase \cite{Koyama:2007ag,Koyama:2007mg,Buchbinder:2007tw,Hinterbichler:2011qk}.  Tolley and Wesley \cite{Tolley:2007nq} analyzed the dynamics of a wide class of contracting cosmological models that generate scale-invariant adiabatic or entropic perturbations and suggested that the problem may be generic.   More specifically, they showed that the cosmological background solutions are not attracted to a fixed point and, from this, concluded that the models are highly sensitive to initial conditions. 

In this paper, we present simple ekpyrotic models that generate a scale-invariant, nearly Gaussian spectrum of density perturbations  but  do {\it not} require fine-tuning of initial conditions.  Although these models belong in the class considered by Tolley and Wesley, we show that  the background solutions are attracted to a fixed-curve along which scale-invariant fluctuations are generated. The existence of such a fixed-curve is sufficient to ensure that the observational predictions are insensitive to the choice of initial conditions. In other words, being attracted to a fixed-point is not necessary to avoid fine-tuning.  

In Section~\ref{sec:secScaling}, we summarize the general argument that suggests the need for finely-tuned initial conditions.  In Section~\ref{sec:secSimple}, we review a simple ekpyrotic model  for which we find a fixed-curved but no fixed-point attractor.  In Section~\ref{sec:secGeneral}, we describe how to construct more general examples that also avoid the need for fine-tuning of initial conditions.  Finally, we discuss the implications for cosmology.  

\section{Scaling solutions, scale-invariance and instability}
\label{sec:secScaling}

{\it Scaling solutions}  are solutions to the equations of motion for which there exists a set of field variables such that all contributions  (treating the kinetic and potential energy densities as distinct)  to the total energy density scale identically with time, keeping their fractional contributions constant.  Scaling background solutions are particularly important because they are exactly solvable and can yield a scale-invariant spectrum of perturbations.  

In Ref.~\cite{Tolley:2007nq}, Tolley and Wesley presented an instability argument that applies to contracting, scaling solutions derived from two-derivative, two-field actions
\begin{eqnarray}
\label{init}
S&=&\int d^4x \sqrt{-g} \frac{1}{2}R\\ \nonumber
& -& \int d^4x \sqrt{-g}\left( \frac{1}{2}G^{ab}(\Phi)g^{\mu\nu}\partial_\mu \Phi_a \partial_\nu \Phi_b-V(\Phi)\right)
\end{eqnarray}
 that possess a continuous symmetry generated by a parameter $\kappa$ such that
 \begin{equation}
\label{conditions}
\frac{d\Phi_a}{d\kappa}=\xi_a(\Phi),\quad g_{\mu\nu} \to e^{\kappa}g_{\mu\nu},\quad S\to e^{\kappa}S. 
\end{equation} 
Here $g_{\mu \nu}$ is the spacetime metric with $(-+++)$ signature convention, $R$ is the Ricci scalar, $\xi_a(\Phi)$ is a function of the two scalar fields $\Phi = \{\Phi_a \}$ where $a=1,2$,  $G^{ab}$ is the metric on field space, and $V(\Phi)$ is the potential energy density; reduced Planck units ($8 \pi \mathrm{G}_N=1$ where $\mathrm{G}_N$ is Newton's gravitational constant) are used throughout.
This symmetry guarantees the existence of a set of field variables $(\Phi_1, \Phi_2) \mapsto (\phi,\sigma)$ such that the Lagrangian density can be rewritten as
\begin{equation}
\label{wtlag}
\mathcal{L}=\frac{1}{2}R-\frac{1}{2}(\partial\sigma)^2-\frac{1}{2} f(\sigma)(\partial\phi)^2-V_0e^{-c\phi}h(\sigma),
\end{equation}
where $c$ is a real constant and $V_0<0$ (that is, the ekpyrotic potential $V(\phi, \sigma)$ is negative).  Along the background solution,   $f(0)=h(0)=1$ and $\sigma=0$.

It proves useful to introduce the dynamical variables 
\begin{eqnarray}
\label{ntconve} 
(w,x,y,z)\equiv\left(\frac{\sqrt{f(\sigma)}\phi'}{\sqrt{6}\mathcal{H}},\frac{\sigma'}{\sqrt{6}\mathcal{H}},-\frac{a\sqrt{-V_0h(\sigma)}e^{-\frac{c}{2}\phi}}{\sqrt{3}\mathcal{H}},\sigma\right)\quad
\end{eqnarray}
where $\tau<0$ is conformal time running from large negative to small negative values, a prime denotes a derivative with respect to
conformal time, $a$ is the scale factor and $\mathcal{H}\equiv a' / a$ is the conformal Hubble parameter. The four variables are dimensionless using reduced Planck units.  With these variables, the Friedmann-Robertson-Walker (FRW) equations of motion become 
\begin{eqnarray}
\label{w}
w,_N &=& 3(w^2+x^2-1)\left(w-\frac{c}{\sqrt{6f(z)}}\right) 
- \sqrt{\frac{3}{2}}\frac{f,_{ z}}{f(z)}xw, \quad
\\
\label{x}
x,_N &=& 3(w^2+x^2-1)\left(x+\frac{1}{\sqrt{6}}\frac{h,_{ z}}{h(z)}\right)+\sqrt{\frac{3}{2}}\frac{f,_{ z}}{f(z)} w^2, \quad
\\
\label{z}
 z,_N & =& \sqrt{6}x.
\end{eqnarray}
Here we introduced the dimensionless time variable $N~\equiv~\ln a$ that denotes the number of $e$-folds of ekpyrotic contraction and runs from large positive to small positive values.
We eliminated $y$ using the Friedmann constraint 
\begin{equation}
\label{fman}
w^2+x^2 - y^2 = 1.
\end{equation} 
The equation of state takes the simple form
\begin{equation}
\label{eosp}
\epsilon = 1-\frac{\mathcal{H}'}{\mathcal{H}^2}=3(w^2+x^2).
\end{equation}
From the Friedmann constraint, we also see that the square of each variable $w,$ $x,$ and $y$ is a fractional contribution to the total energy density: $w^2$ is the $\phi$-kinetic energy; $x^2$ is the $\sigma$-kinetic energy; and $y^2$ is the potential energy. The equation-of-state parameter $\epsilon$ is the sum of kinetic energies.

As can be verified by direct substitution, the background Eqs.~(\ref{w}-\ref{z}) admit a fixed-point solution,
\begin{equation}
\label{wtfp}
\left(w, x, y, z\right)=\left(\frac{c}{\sqrt{6}},0 , \sqrt{\frac{c^2}{6}-1} , 0\right),
\end{equation}
in which $\sigma=0$ with  and only $\phi$ is changing, provided the constraint $h,_\sigma(0)=-c^2 f,_\sigma(0)/(c^2-6)$.  Obviously, this is a scaling solution since any fractional contribution to the total energy density, $w^2,$ $x^2,$ and $y^2$, is constant.

The cosmological background solutions correspond to a field-space trajectory like the one shown in Fig.~\ref{field}. Perturbations of this trajectory can be decomposed into those along the red curve (\emph{adiabatic} perturbations)  and  perpendicular to it (\emph{entropic} perturbations). 
 
\begin{figure}\label{field}
\centering
\includegraphics[scale=0.75]{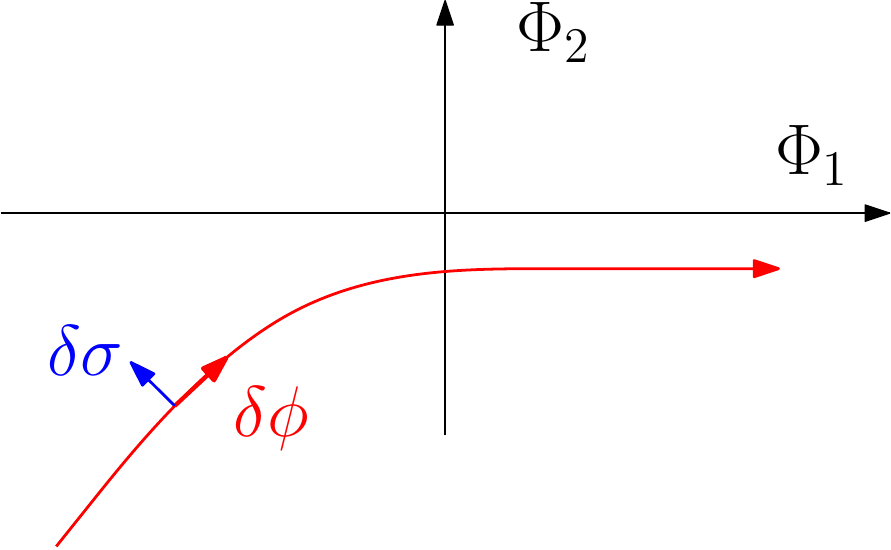}
\caption{\label{field} A schematic trajectory of a solution through field-space:  The field variables $(\phi,\sigma)$ were constructed so that, along the field-space trajectory in Eq.~\eqref{wtfp}, only $\phi$ varies and $\sigma=0$.  Perturbations $\delta\phi$ are tangent to the trajectory and are  adiabatic; perturbations $\delta \sigma$ are orthogonal to the trajectory and are entropic.}
\end{figure}
The instability argument connects the stability of the solution in Eq.~\eqref{wtfp} with the spectral indices of its perturbations.  The basic idea makes use of the fact that, since  $z\equiv \sigma=0$ in the background solution, the second order action and, hence, the perturbation spectra derived from it are determined by a few parameters
\begin{equation}
\label{parameters}
 \{ c\,; f,_\sigma(0); f,_{\sigma\sigma}(0); h,_{\sigma\sigma}(0) \}.
\end{equation}
Linearized around the fixed-point in Eq.~\eqref{wtfp}, the background Eqs.~(\ref{w}-\ref{z}) reduce to a matrix equation, where the parameters in Eq.~\eqref{parameters} determine the eigenvalues of the matrix.
Tolley and Wesley's analysis showed that any combination of the parameters in Eq.~\eqref{parameters} that results in a scale-invariant spectrum of perturbations (entropic or adiabatic) renders the background solution dynamically unstable to perturbations in the sense that the matrix associated with the linearized system has at least one negative eigenvalue. In a contracting universe, a negative eigenvalue means a dynamically unstable direction in ($w$-$x$-$z$)-space.  

\begin{figure}
\centering
\includegraphics[scale=0.45]{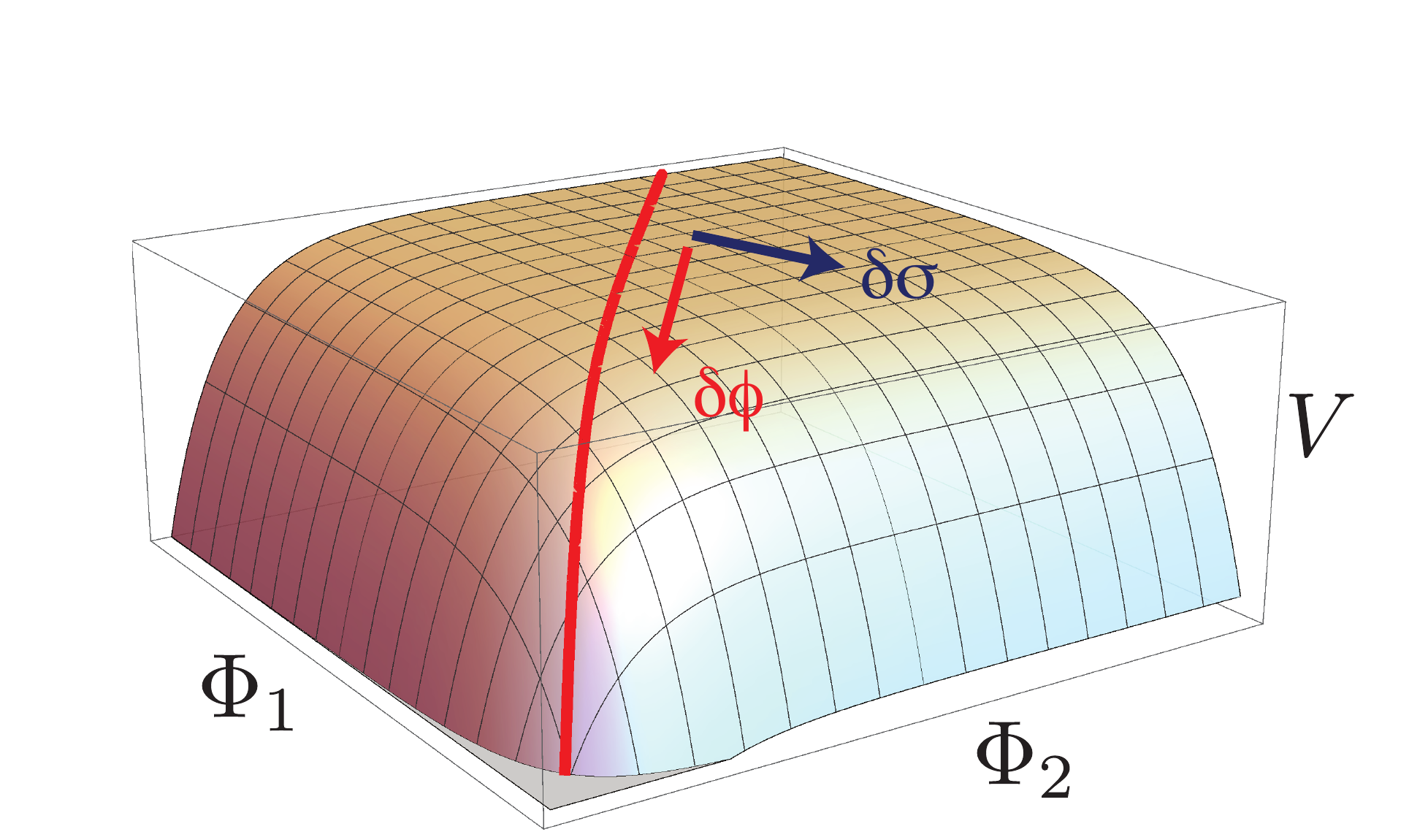}
\caption{\label{pot} The instability of the scaling solution corresponding to the Lagrangian density in Eq.~\eqref{doubexp}:  The background cosmological solution corresponds to a trajectory along the ridge of the potential as indicated by the red arrow and red curve.  Quantum fluctuations along the trajectory produce a blue spectrum of adiabatic perturbations and fluctuations normal to the trajectory produce a scale-invariant  spectrum of entropic perturbations.  After the ekpyrotic phase, the entropic perturbations convert into adiabatic perturbations due to the bending of the trajectory curve (not shown here).  The fact that the background trajectory is a ridge means that it is unstable if the initial conditions are sufficiently far from the ridge.}
\end{figure}

An example is given by the Lagrangian density
\begin{equation}
\label{doubexp}
\mathcal{L}=\frac{1}{2}R-\frac{1}{2}(\partial\Phi_1)^2+\frac{1}{2}(\partial\Phi_2)^2-\tilde{V}_0e^{-c_1\Phi_1}-\tilde{V}_0e^{-c_2 \Phi_2},
\end{equation}
where $c_1, c_2$ are positive-definite constants and $\tilde{V}_0<0$.
The corresponding FRW equations of motion admit a scaling solution with $\Phi_i=A_i \ln |\tau| + B_i$ and $c_1A_1=c_2A_2$ that has been shown to generate a scale-invariant spectrum of entropic perturbations \cite{Lehners:2007ac, Lehners:2008vx}.  According to the instability argument, it should have an unstable background which in this simple case ($G_{ab}=\delta_{ab}$) can be depicted as in Fig. \ref{pot}.  

For the purpose of illustration, we briefly outline how the instability emerges from a negative eigenvalue of the linearized system. The change of variables, 
\begin{eqnarray}
\phi &=& \frac{c_2 \Phi_1+c_1\Phi_2}{\sqrt{c_1^2+c_2^2}},
\\
\sigma &=& \frac{c_1\Phi_1-c_2\Phi_2}{\sqrt{c_1^2+c_2^2}}+\sigma_0
\end{eqnarray}
with $\sigma_0=2 \ln (c_2/c_1)/ (c_1^2+c_2^2) $, brings the Lagrangian density in Eq.~\eqref{doubexp} to the form of Eq.~\eqref{wtlag}.  The coupling function to the kinetic energy of $\phi$, $f(\sigma)$, and the coupling function to the potential energy of $\phi$, $h(\sigma)$, are given by
\begin{eqnarray}
f(\sigma)&=&1,\\
h(\sigma)&=&1+\frac{c^2}{2}\sigma^2+\mathcal{O}(\sigma^3),
\end{eqnarray}
and the parameters, $c$ and $V_0$ are defined such that
\begin{eqnarray}
\frac{1}{c^2} & = &  \frac{1}{c_1^2} + \frac{1}{c_2^2} ,
\\
V_0 &=& \left( \left(\frac{c_2}{c_1}\right)^{ \frac{2 c_1^2 }{ c_1^2+c_2^2} } + \left(\frac{c_1}{c_2}\right)^{\frac{2c_2^2}{c_1^2+c_2^2}}\right)\tilde{V}_0.
\end{eqnarray}
Linearizing the background Eqs.~(\ref{w}-\ref{z}) about the fixed-point in Eq.~\eqref{wtfp}, the perturbations $(\delta w,\delta x, \delta z)\equiv (w-\frac{c}{\sqrt{6}}, x,z)$ satisfy
\begin{equation}
\label{linear}
\left(\begin{array}{c}
\delta w,_N\\
\delta x,_N\\
\delta z,_N
\end{array}\right)=M \cdot \left(\begin{array}{c}
\delta w\\
\delta x\\
\delta z
\end{array}\right)
\end{equation}
with $M$ defined as
\begin{equation}
M\equiv
\left(
\begin{array}{ccc}
 \frac{1}{2} \left(c^2-6\right) & 0 & 0 \\
 0 & \frac{1}{2} \left(c^2-6\right) & \frac{c^2}{2\sqrt{6}} \left(c^2-6\right) \\
 0 & \sqrt{6} & 0 \\
\end{array}
\right){.}
\end{equation}
The eigenvalues of $M$ are $\left(c^2-6\right)/2$ and $\left( (c^2 - 6) \pm \sqrt{9c^4 - 60 c^2 + 36}\right)/4$. Note that the smallest eigenvalue is negative for ekpyrosis, as is clear from substituting Eq.~\eqref{wtfp} into the equation of state, Eq.~\eqref{eosp}:  $\epsilon > 3$ requires $c > \sqrt{6}$.  Therefore, as the universe contracts, $N$ decreases, and perturbations along the eigenvector corresponding to the negative eigenvalue grow so that the system is carried away from the fixed-point solution in Eq.~\eqref{wtfp}.   In this case, the negative eigenvalue means that the initial conditions for the fields must be fine-tuned to lie close to the trajectory or else the fields will evolve far-off course as illustrated in Fig.~\ref{pot}.     
   
\section{Ekpyrosis and Scale-invariance without fine tuning}
\label{sec:secSimple}
  
In this section, we describe the case where the negative eigenvalue exists but is physically irrelevant. As we will show, the occurrence of the negative eigenvalue only means that the attractor is a fixed-curve rather than a fixed point. 

We consider the Lagrangian density first discussed by Li in Ref.~\cite{Li:2014qwa}, 
\begin{equation}
\label{lilag}
\mathcal{L}=\frac{1}{2}R-\frac{1}{2}(\partial\psi)^2-\frac{1}{2}e^{-\lambda\psi}(\partial\chi)^2-\tilde{V}_0e^{-\lambda \psi},
\end{equation}
where $\lambda$ is a positive and $\tilde{V}_0<0$.
The model involves an ekpyrotic field, $\psi$, with a negative potential, similar to ordinary, single-field ekpyrosis.  The novel feature is the non-canonical, exponential coupling to the massless spectator field, $\chi$.  
We begin with the simple case where $V(\psi)=0$.   This corresponds to the borderline ekpyrotic equation of state $\epsilon = 3$. Then, we generalize to $V(\psi) \neq 0$ ($\epsilon>3$) and provide a full, analytic treatment.

\subsection{$V(\psi) =0$}
\begin{figure}
\includegraphics[scale=.4]{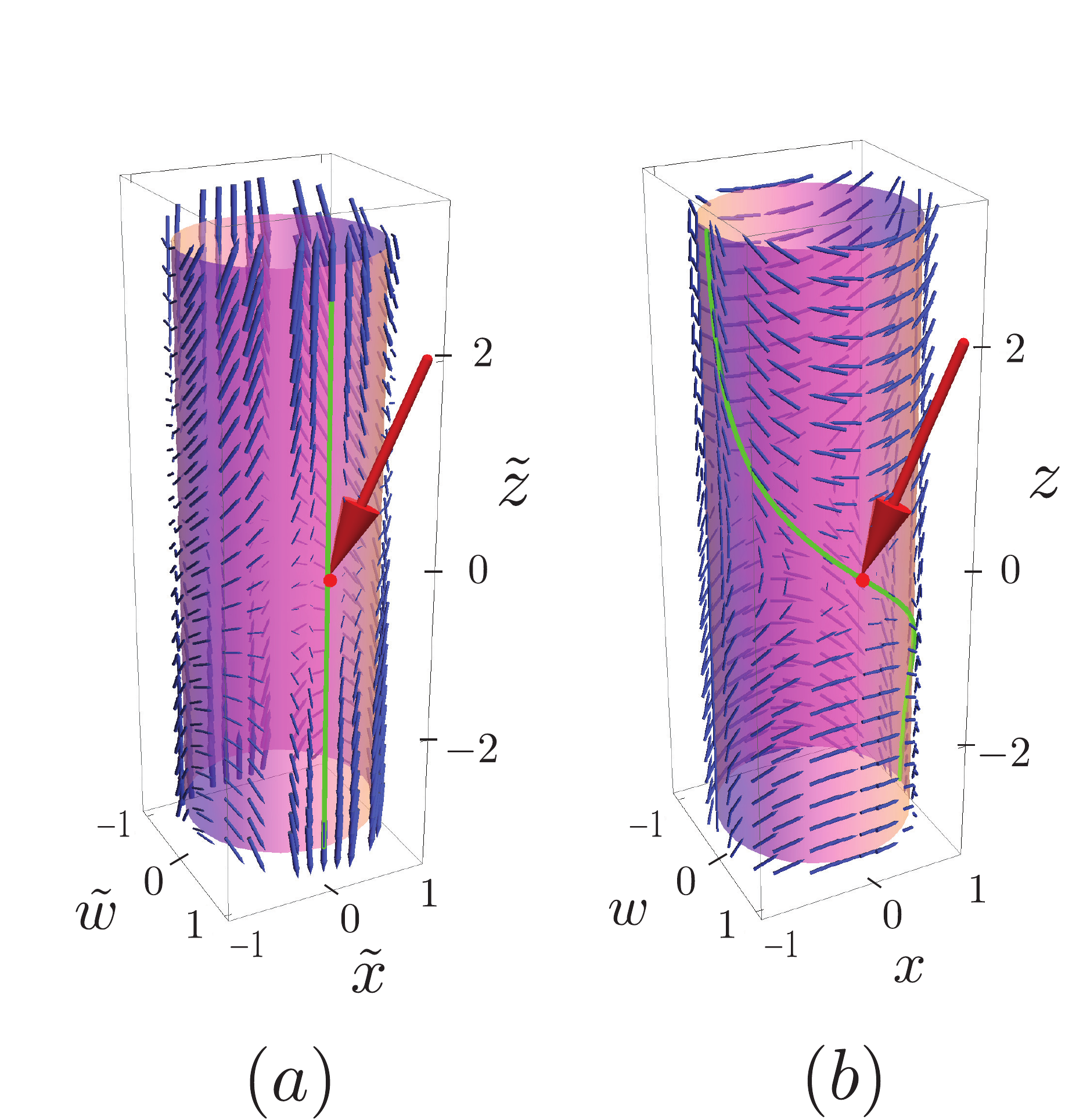}
\caption{\label{twphasespace} 
The Lagrangian density in Eq.~\eqref{lilag} with $V=0$  and $\lambda = \sqrt{6}$:  Green curves and red arrows map onto each other according to the transformations in Eqs.~(\ref{s1xrw}-\ref{s1xrz}). Solutions follow the blue arrows. 
FIG.~3.a.  {The trajectories in $(\psi,\chi)$ variables:}
For $\tilde{y}=0, \;
(\tilde{w},\, \tilde{x},\, \tilde{z}) = \left( \psi' / (\sqrt{6}\mathcal{H}), \; e^{- \frac{\lambda}{2} \psi } \chi' / (\sqrt{6}\mathcal{H}), \; e^{- \frac{\lambda}{2} \psi } (\chi-\chi_0)\right).$ The green curve, $(\tilde{w},\tilde{x},\tilde{z})=(1,0,\tilde{z})$, is the fixed-curve attractor. The red arrow points to the  fixed-point $(\tilde{w},\tilde{x},\tilde{z})=(1,0,0).$ 
FIG.~3.b. The trajectories in $(\phi,\sigma)$ variables:
For $y=0,\; (w,\, x,\, z) = \left( \sqrt{f(\sigma)}\, \phi' / (\sqrt{6}\,\mathcal{H}), \; \sigma' / (\sqrt{6}\,\mathcal{H}), \;\sigma \right).$ The green curve, $(w,x,z)=\left(\text{sech}(\lambda\, z/2), -\tanh( \lambda\, z/2) ,z\right)$, is the fixed-curve attractor.  The red arrow points to the fixed-point $(w,x,z)=(1,0,0)$.}
\end{figure}

The basic idea is captured in Fig. \ref{twphasespace}a, which  illustrates background trajectories corresponding to different initial conditions in the space $(\tilde{w},\tilde{x},\cancelto{0}{\tilde{y}},\tilde{z})=
$
\begin{equation}
\label{lisphsp}
\left(
\frac{\psi'}{\sqrt{6}\mathcal{H}}, \frac{e^{-\frac{\lambda}{2}\psi}\chi'}{\sqrt{6}\mathcal{H}}, \cancelto{0}{\frac{- a\sqrt{-\tilde{V}_0}e^{-\frac{\lambda}{2}\psi}}{\sqrt{3}\mathcal{H}}},e^{-\frac{\lambda}{2}\psi}(\chi-\chi_0)\right).
\end{equation} 
We use the superscript $\sim$ to distinguish quantities expressed in $(\psi,\chi)$ variables from those expressed in the special $(\phi,\sigma)$ variables that were used to derive the original instability argument.

Any set of initial conditions for $a, \psi, \psi', \chi$, and $ \chi'$ corresponds to a particular point on the cylinder.  The background solution follows the blue arrow originating at this point. There are two special initial conditions at the points $(\tilde{w},\tilde{x},\tilde{z})=(\pm 1,0,0)$; the one with the $+$ sign corresponds to the red arrow in Fig.~\ref{twphasespace}a.  These two points are special because the blue arrows vanish here, {\it i.e.,}  if the background solution starts here, it stays here.  Hence, these are fixed-point solutions.  

For any other initial conditions, ({\it i.e.,} any other point on the cylinder), the blue arrows carry the background solution toward the green curve.  The green curve is therefore a strong attractor for generic initial conditions.  Henceforth, we call such an attractor curve a ``fixed-curve attractor.'' Solutions evolving along this curve are scaling solutions since $\tilde{w}, \tilde{x},$ and $\tilde{y}$ are constant.  They describe a universe dominated by the kinetic energy of the adiabatic field, $\psi$; the entropic field, $\chi$, is constant.  
For the case ($\lambda=\sqrt{6}$) depicted in Fig.~\ref{twphasespace}, these solutions can be shown to generate a scale-invariant spectrum of perturbations in the entropic field, $\chi$.  

As for the negative eigenvalue associated with the linearized equations of motion, in the cases discussed in this paper, it only indicates the existence of a fixed-curve attractor instead of a fixed-point attractor.  The existence of a fixed-curve attractor means there is no need for fine-tuning of initial conditions.

One can describe the same dynamics in the special $(\phi,\sigma)$ variables (see Fig. \ref{twphasespace}b). Since the Lagrangian density in Eq.~\eqref{lilag} has the shift symmetry of Eq.~\eqref{conditions} assumed by the instability argument, it can be put into the form of Eq.~\eqref{wtlag} through a variable transformation $(\psi,\chi)\mapsto(\phi,\sigma)$,
\begin{eqnarray}
\label{coord1}
\psi &=& \phi+\frac{2}{\lambda}\ln\left[\text{sech}\left(\frac{\lambda\sigma}{2}\right)\right]+\psi_0,\\
\label{coord2}
\chi &=& \frac{2}{\lambda}e^{\frac{\lambda}{2}(\phi+\psi_0)}\tanh\left(\frac{\lambda\sigma}{2}\right)+\chi_0,
\end{eqnarray}
where $\psi_0$ is a real constant.  Substituting this transformation into Eq.~\eqref{lilag} yields the Lagrangian density in Eq.~\eqref{wtlag} with $f(\sigma)=h(\sigma)=\cosh^2( \lambda\sigma/2)$ and $c=\lambda$.
The two Lagrangian densities describe the same theory in different field variables:  the cylinder on the right is a twisted version of the one on the left.  It is clear from the blue arrows that the green curve in Fig.~\ref{twphasespace}b is an attractor, just like the green curve in Fig. \ref{twphasespace}a.  Solutions along the green-curve attractor generate a scale-invariant spectrum of entropic perturbations.  

\subsection{$V(\psi) \neq 0$}
We extend our analysis to the more general form of the Lagrangian density in Eq.~\eqref{lilag} that applies both to $V(\psi)=0$  and to $V(\psi)\neq 0$.  

Extremizing the action with respect to variations of the fields $(\psi,\chi)$ yields the field equations
\begin{eqnarray}
&& \psi''+2\mathcal{H}\psi' - \lambda \tilde{V}_0e^{-\lambda\psi}a^2+\frac{\lambda}{2}e^{-\lambda\psi}\chi'^2=0, 
\\
&& \chi'' +2\mathcal{H}\chi'-\lambda \psi'\chi'=0.
\end{eqnarray}
The Friedmann constraint is 
\begin{equation}
\mathcal{H}^2=\frac{1}{6}\left(\psi'^2+e^{-\lambda\psi}\chi'^2+2a^2\tilde{V}_0e^{-\lambda\psi}\right).
\end{equation}
Using the variables defined in Eq.~\eqref{lisphsp}, these equations can be recast as the autonomous, dynamical system
\begin{eqnarray}
\label{liw}
\tilde{w},_N &=& 3(\tilde{w}^2+\tilde{x}^2-1)\left(\tilde{w}-\frac{\lambda}{\sqrt{6}}\right)-\sqrt{\frac{3}{2}}\lambda \tilde{x}^2,
\\
\label{lix}
\tilde{x},_N &=& 3\tilde{x}(\tilde{w}^2+\tilde{x}^2-1)+\sqrt{\frac{3}{2}}\lambda \tilde{w}\tilde{x},
\\
\label{liz}
\tilde{z},_N &=& -\sqrt{\frac{3}{2}}\lambda \tilde{w} \tilde{z} +\sqrt{6} \tilde{x}.
\end{eqnarray}

If $V\neq0$, this system admits three fixed-point solutions at $(\tilde{w},\tilde{x},\tilde{y},\tilde{z})=$ 
\begin{eqnarray}
&& \left(-1,0,0,0\right)
\label{notredarrow}, \\
&& \left(+1,0,0,0\right)
\label{redarrow},\\
&& \left(\frac{\lambda}{\sqrt{6}},0,\sqrt{\frac{\lambda^2}{6}-1},0\right)\label{imp},
\end{eqnarray}
all of which are unstable ({\it i.e.,} associated with a negative eigenvalue).  The third fixed-point solution given by Eq.~\eqref{imp} bisects two fixed-\emph{curve} solutions
\begin{equation}
\label{lison}
\left(\tilde{w},\tilde{x},\tilde{y},\tilde{z}\right)=\left(\frac{\lambda}{\sqrt{6}},0,\sqrt{\frac{\lambda^2}{6}-1},\pm \tilde{Z}\right)
\end{equation}
with $\tilde{Z} \propto e^{-\frac{\lambda^2}{2} N}$ that generate a scale-invariant spectrum of entropic perturbations, as shown in Ref.~\cite{Li:2014qwa}.   

If $V=\tilde{y}=0$, $\lambda^2$ must be $6$ in order for the fixed-point in Eq.~\eqref{imp} to be a solution.  For $\lambda=\sqrt{6}$, this coincides with the fixed-point in Eq.~\eqref{redarrow} and corresponds to the red arrow shown in Fig.~\ref{twphasespace}a;  Eq.~\eqref{lison} parameterizes the vertical, green, fixed-curve attractor.

Changing variables $(\psi,\chi) \mapsto (\phi,\sigma)$ as defined in Eqs.~\eqref{coord1} and \eqref{coord2},  the Lagrangian density in Eq.~\eqref{lilag} takes the form of Eq.~\eqref{wtlag} with $f(\sigma)=h(\sigma)=\cosh^2\left( \lambda \sigma/2 \right), c=\lambda,$ and $V_0=\tilde{V}_0e^{-\lambda\psi_0}.$
Repeating the same analysis in the new variables defined in Eq.~\eqref{ntconve}, the equations of motion become
\begin{eqnarray}
\label{wtw}
w,_N &=& 3(w^2+x^2-1)\left(w-\frac{\lambda}{\sqrt{6}}\text{sech}\left(\lambda z \right)\right)
\nonumber \\
&-& \sqrt{6}c \tanh(\lambda z)\, x w,
\\
\label{wtx}
x,_N &=& 3(w^2+x^2-1)\left(x+\sqrt{\frac{2}{3}} \lambda \tanh (\lambda z)\right)
\nonumber \\
&+& \sqrt{6}c \tanh(\lambda z)\, w^2,
\\
\label{wtz}
z,_N &=& \sqrt{6}x.
\end{eqnarray}

The variable transformations as defined in Eqs.~\eqref{coord1} and \eqref{coord2} imply the following relationship between the variables $(w,x,y,z)\mapsto(\tilde{w},\tilde{x},\tilde{y},\tilde{z})$:
 \begin{eqnarray}
\label{s1xrw}
\tilde{w} &=& \text{sech}\left(\frac{\lambda}{2}z\right)w-\tanh\left(\frac{\lambda}{2}z\right)x,
\\
\label{s1xrx}
\tilde{x} &=& \tanh\left(\frac{\lambda}{2}z\right)w+\text{sech}\left(\frac{\lambda}{2}z\right)x,
\\
\label{s1xry}
\tilde{y} &=& y,
\\
\label{s1xrz}
\tilde{z} &=& \frac{2}{\lambda}\sinh\left(\frac{\lambda}{2}z\right).  
\end{eqnarray}
If $V=0$, these transformations  quantify how to ``twist'' Fig.~\ref{twphasespace}b to generate Fig. \ref{twphasespace}a.

The fixed-point solutions in Eqs.~(\ref{notredarrow}-\ref{imp}) are given in the new variables $(w,x,y,z)$ as
\begin{eqnarray}
&&\left(-1,0,0,0\right),
\label{notredarrow2}\\
&& \left(+1,0,0,0\right),
\label{redarrow2}\\
&& \left(\frac{\lambda}{\sqrt{6}},0,\sqrt{\frac{\lambda^2}{6}-1},0\right),\label{imp2}
\end{eqnarray}
and the fixed-curve solutions in Eq.~\eqref{lison} are 
\begin{equation}
\label{TW-A}
\left(\frac{\lambda}{\sqrt{6}}\text{sech}\left(\frac{\lambda}{2}z\right),-\frac{\lambda}{\sqrt{6}}\tanh\left(\frac{\lambda}{2}z\right), \sqrt{\frac{\lambda^2}{6}-1},\pm Z\right)
\end{equation}
with $Z = ( 2 / \lambda) \sinh^{-1}\left( \lambda\, \tilde{Z}/2 \right)$.
These fixed-curves lie on the surface of the cylinder $w^2+x^2= \lambda^2 /6$.  
For $V= 0$ and $\lambda=\sqrt{6}$, Eq.~\eqref{TW-A} corresponds to the twisted green curve that is confined to the surface of the unit cylinder in Fig. \ref{twphasespace}b.    

Direct substitution verifies that the curves in Eq.~\eqref{lison} and Eq.~\eqref{TW-A} are solutions to the background equations given in Eqs.~(\ref{liw}-\ref{liz}) and Eqs.~(\ref{wtw}-\ref{wtz}) for both $V=0$ with $\lambda=\sqrt{6}$ and for $V\neq 0$.

To show the existence of a negative eigenvalue, we linearize the equations of motion about the fixed-points, in Eq.~\eqref{imp} and \eqref{imp2}, respectively, for the two sets of variables.   Linearizing Eqs.~(\ref{liw}-\ref{liz}) about Eq.~\eqref{imp} yields a matrix equation like that given in Eq.~\eqref{linear} with 
$\left( 
\delta \tilde{w}, \delta{\tilde{x}}, \delta{\tilde{z}}\right) = \left(\tilde{w} - \lambda/\sqrt{6}, \tilde{x}, \tilde{z}
\right)$ 
and 
\begin{equation}
\tilde{M}\equiv
\left(
\begin{array}{ccc}
 \frac{\lambda ^2}{2}-3 & 0 & 0 \\
 0 & \lambda ^2-3 & 0 \\
 0 & \sqrt{6} & -\frac{\lambda ^2}{2} \\
\end{array}
\right).
\end{equation}
Similarly, linearizing Eqs.~(\ref{wtx}-\ref{wtz}) about Eq.~\eqref{imp2} yields Eq.~\eqref{linear} with $\left(\delta w,\delta{x},\delta{z}\right) = \left(w - \lambda/\sqrt{6}, x, z \right)$ and 
\begin{equation}
M\equiv
\left(
\begin{array}{ccc}
 \frac{\lambda^2}{2}-3 & 0 & 0 \\
 0 &  \frac{\lambda^2}{2}-3 & \frac{\lambda^2 \left(\lambda^2-3\right)}{2 \sqrt{6}} \\
 0 & \sqrt{6} & 0 \\
\end{array}
\right).
\end{equation}
Both $\tilde{M}$ and $M$  have eigenvalues $\left \{- \lambda^2 / 2, \lambda^2 /2 - 3,\lambda^2-3\right \}$, the first of which is negative. In the first set of variables, the eigenvector corresponding to the eigenvalue $-\lambda^2/2$ is parallel to the unit vector in the $\tilde{z}$-direction, which is tangent to the fixed-curve solution in Eq.~\eqref{lison}.  In the second set of variables, the eigenvector associated with the eigenvalue $-\lambda^2/2$ is parallel to a linear combination of unit vectors $\hat{x}, \hat{z}$, namely $\hat{z} - \lambda^2/(2\sqrt{6})\, \hat{x}$, that is tangent to the fixed-curve solution in Eq.~\eqref{TW-A}.  

For the case, $V=0$ and $\lambda=\sqrt{6}$, these eigenvectors are tangent to the green fixed-curves in Fig.~\ref{twphasespace} at the red arrows. The existence of a negative eigenvalue in this model is harmless, since it only means that the system is attracted to a fixed-curve solution (instead of a fixed-point solution) that generates a scale-invariant spectrum of entropic perturbations. 

\subsection{Further Generalizations}
Although the remainder of this work will consider actions with the shift symmetry in Eq.~\eqref{conditions}, our results can be generalized to cases without shift symmetry.  
For example, the field contribution to the shift-symmetric Lagrangian density in Eq.~\eqref{lilag} is a special case of the more general Lagrangian density
\begin{equation}
\label{newlag3}
\mathcal{L}=\frac{1}{2}R-\frac{1}{2}(\partial\psi)^2-\frac{1}{2}e^{-\lambda\psi}(\partial\chi)^2 - \left(1+r(\chi)\right) V_0e^{-\mu\psi} + q(\chi)
\end{equation} 
with $\mu=\lambda$ and $r(\chi)=q(\chi)=0$.  The addition of $q(\chi)$ and $r(\chi)$ breaks the shift symmetry since
\begin{eqnarray}
V &=& \left( 1+r(\chi)\right)  V_0e^{-\mu\psi} + q(\chi)  \nonumber
\\
& \to & \left(1+r(e^{\frac{\lambda}{2\mu}\kappa}\chi) \right) V_0e^{-\mu\psi-\kappa} + r(e^{\frac{\lambda}{2\mu}\kappa}\chi) \nonumber
\\
&\neq& e^{-\kappa}V.
\end{eqnarray}  
If  $\mu= \lambda$, the ekpyrotic Lagrangian density in Eq.~\eqref{newlag3} admits a scaling solution that is a fixed-curve attractor with $\chi'=0$ and that generates a scale-invariant spectrum of entropic perturbations. This is due to the fact that, as  $\chi' \to 0$, $r(\chi)$ and $q(\chi)$ approach constants $r(\chi_0)$ and $q(\chi_0)$. The first, $r(\chi_0)$, can be reabsorbed into $V_0$, and the second, $q(\chi_0)$, is negligible along the fixed-curved attractor.

\section{Constructing New Models}
\label{sec:secGeneral}

In this section, we derive the most general ekpyrotic, two-field Lagrangian density with shift symmetry  that admits scaling solutions which are either fixed-point or fixed-curved attractors and generate a scale-invariant spectrum of entropy perturbations.  

First, we consider the Lagrangian density in Eq.~\eqref{wtlag} with arbitrary parameters and couplings,
\begin{equation}
\label{choice}
\{V_0<0, c\in \mathbb{R}, h(\sigma)>0, f(\sigma)>0\}.
\end{equation}
In the Appendix, we show that the combined conditions of shift symmetry, scaling solution, fixed-curved attractor and scale-invariant spectrum of perturbations imply the following properties:
\begin{enumerate} 
\item[P1:]  $\underset{|\sigma|\to\infty}{\lim} f(\sigma) = \infty$  monotonically;
\item[P2:]
$\underset{|\sigma|\to\infty}{\lim} h(\sigma)\propto e^{-\mu\sigma}$; 
\item[P3:]
$\underset{|\sigma|\to\infty}{\lim} (w,x,y,z) = \left(0, \frac{\mu}{\sqrt{6}}, \sqrt{ \frac{\mu^2}{6} -1}, -\text{sgn}(\mu) \infty \right)$;
\item[P4:]
$|\mu|>\sqrt{6}$.
\end{enumerate}
Property~P1 says that the coupling $f(\sigma)$ must grow without bound.  Property~P2 constrains the form of the potential energy density to be exponential at late times. Property~P3 defines the scaling solution. It implies $w=0$ so that the $\phi$ field is fixed; furthermore, since the background equations in Eqs.~(\ref{w}-\ref{z}) depend explicitly on $\sigma$ which will vary, it also implies that the scaling solution is a fixed-curve (rather than a fixed-point) in ($w$-$x$-$z$)-space. Property~P4 is necessary for ekpyrosis ($\epsilon>3$) as follows from substituting this solution into the equation of state, Eq.~\eqref{eosp}.  

The example from the last section has these four properties.  At late times, the fixed-curve attractor in Eq.~\eqref{TW-A} goes to $z\equiv\sigma \to\pm\infty$. In this limit, $f(\sigma)=h(\sigma)=\cosh^2(\lambda \sigma/2 )$ is dominated by the single exponential $e^{|\lambda\sigma|}/4$.  For example, if $\lambda>0$ and $\sigma\to-\infty$, choosing $\mu=\lambda$ in the solution in property~P3 reproduces Eq.~\eqref{TW-A} at late times. Similar arguments apply for the different combinations of $\text{sgn}(\lambda)$ and $\text{sgn}(\sigma)$.

Assuming properties P1 thru P4, the only remaining degrees of freedom are the parameters, $c, V_0$, and the late-time behavior of $f(\sigma)$, modulo property~P1.  We show now that, given these four properties, it is possible to obtain a scale-invariant spectrum for the entropic modes but not for the adiabatic modes.

We perturb Einstein's equations about the fixed-curve solution specified by Property~P3, working in the longitudinal gauge \cite{Bardeen:1980kt,Bardeen:1983qw} where the metric takes the form 
\begin{equation}
\text{d}s^2=a^2\left(-(1+2\Phi)\text{d}\tau^2+(1-2\Phi)\text{d}\vec{x}^2\right).
\end{equation}
Since $\phi'=0$ along the background solution in property~P3, the quantity $Q_s \equiv \sqrt{f(\sigma)}\delta\phi$ is automatically gauge-invariant and represents the entropy perturbation; the Mukhanov-Sasaki variable $Q_\sigma \equiv \delta\sigma + (\sigma'/\mathcal{H})\,\Phi$ is also gauge-invariant and represents the adiabatic perturbation \cite{Sasaki:1986hm,Mukhanov:1988jd}.  Property~P3 implies that the equation of state $\epsilon=\mu^2/2$ and, therefore,  the conformal Hubble parameter is $\mathcal{H}^{-1}=(1-\epsilon)(-\tau)<0$.   Then, the mode functions $u_\sigma \equiv a\, Q_\sigma$ and $u_s\equiv a\, Q_s$ can be shown to satisfy 
\begin{eqnarray}
\label{modefunctionsa}
u_\sigma''+\left(k^2 - \frac{\theta^\sigma}{(-\tau)^2} \right)u_\sigma &=&\frac{\alpha}{(-\tau)^2} u_s + \frac{\beta}{(-\tau)} u_s',\\
\label{modefunctionse}
u_s''+\left( k^2 - \frac{\theta^s}{(-\tau)^2}  \right)u_s & = & \frac{\gamma}{(-\tau)^2} u_\sigma + \frac{\delta}{(-\tau)} u_\sigma',
\end{eqnarray}
where the the background-dependent quantities can be derived, for example, from the expressions in Ref.~\cite{[{}] [{.  In their notation, $V_{\sigma} = \mu V$, $V_{s}= (c/\sqrt{f})V$, $V_{\sigma \sigma} = \mu^2 V$, $V_{s s}= (c^2/f)V$, $V_{\sigma s}=(\mu c/\sqrt{f})V$, and $a^2V = - (\mu^2-6)\mathcal{H}^2/2= -2 (-\tau)^{-2}(\mu^2-6)/(\mu^2-2)^2 $  where all of these quantities are negative.}] Lalak:2007vi}: 
\begin{eqnarray}
\label{constanttermsthsig}
 \theta^\sigma  & = & -   2 \frac{\mu ^2-4}{(\mu ^2-2)^2 } -
\frac{ \left(\mu ^2-6\right)^2}{\mu^2 (\mu ^2-2)^2 }  \frac{c^2}{f},
\\
%%%%%
\label{constanttermsths}
\theta^s & = & 
- 2 \frac{\mu^2 - 4}{(\mu^2-2)^2} 
+ 3 \frac{\mu^2-6}{\mu^2(\mu^2-2)} \frac{c^2}{f}
-  \frac{\mu(\mu^2-6)}{(\mu^2-2)^2} \frac{f,_\sigma}{f }\qquad
\nonumber\\
&& - \frac{\mu^2}{(\mu^2-2)^2}  \left(\frac{f,_\sigma}{f}\right)^2
+ \frac{2\mu^2}{(\mu^2-2)^2}\frac{f,_{\sigma\sigma}}{f} ,
\\
\label{constanttermsa}
\alpha & = & 4 \frac{\mu^2 - 6}{\mu (\mu^2 - 2 )^2} \frac{c}{ \sqrt{f} } 
+ 2 \frac{\mu^2 - 6}{(\mu^2 - 2)^2} \frac{c}{\sqrt{f}} \frac{f,_\sigma}{f}
,\quad
\\
\label{constanttermsb}
\beta & = & 2 \frac{\mu ^2 - 6 }{\mu  \left(\mu ^2-2\right)} \frac{c}{  \sqrt{f }},
\\
\label{constanttermsg}
\gamma & = & - 4\frac{ \mu ^2 - 6 }{\mu  \left(\mu ^2-2\right)^2 } \frac{c}{ \sqrt{f}},
\\
\label{constanttermsd}
\delta & = & -\beta .
\end{eqnarray}
These variables are generally time-dependent because $f=f(\sigma)$ is a function of $\tau$. 
Substituting the expression for $\mathcal{H}^{-1}$ into the definition of $x$ given in Eq.~(\ref{ntconve}), we find
\begin{equation}
f,_{\sigma} = \frac{f'}{ \sigma'}= - f'\, \frac{\mu^2-2}{2 \mu} (-\tau). 
\end{equation}
Using this expression, Eqs.~(\ref{constanttermsthsig}-\ref{constanttermsd}) can be rewritten as:
\begin{eqnarray}
\label{constantterms2thsig}
 \theta^\sigma  & = & -   2 \frac{\mu ^2-4}{(\mu ^2-2)^2 } -
\frac{ \left(\mu ^2-6\right)^2}{\mu^2 (\mu ^2-2)^2 }  \frac{c^2}{f},
\\
\label{constantterms2ths}
\theta^s & = & 
-2 \frac{ \mu^2-4 }{(\mu^2-2)^2}  
+ 3 \frac{\mu^2-6}{\mu^2(\mu^2-2)} \frac{c^2}{f}
- \frac{2}{\mu^2-2} \frac{f'}{ f} (-\tau)
\nonumber \\
&&-  \frac{1}{4}\left(\frac{f' }{f} (- \tau) \right)^2
+\frac{1}{2} \frac{f''}{f}(-\tau)^2 ,\quad
\\
\label{constanttermst2a}
\alpha & = & 4 \frac{\mu^2 - 6}{\mu (\mu^2 - 2 )^2} \frac{c}{ \sqrt{f} } 
+ \frac{\mu^2 - 6}{\mu (\mu^2 - 2)} \frac{c}{\sqrt{f}} \frac{f'}{ f} (-\tau) ,
\\
\label{constantterms2b}
\beta & = & 2 \frac{\mu ^2 - 6 }{\mu  \left(\mu ^2-2\right)} \frac{c}{  \sqrt{f }},
\\
\label{constantterms2g}
\gamma & = & - 4\frac{ \mu ^2 - 6 }{\mu  \left(\mu ^2-2\right)^2 } \frac{c}{ \sqrt{f}},
\\
\label{constantterms2d}
\delta & = & -\beta.
\end{eqnarray}

Eqs.~\eqref{modefunctionsa} and \eqref{modefunctionse} are a coupled, linear system of differential equations which must be solved as $\tau \to 0^-$ to find the spectra.  Depending on the growth rate of the coupling function $f$, different terms in Eqs.~(\ref{constantterms2thsig}-\ref{constantterms2d}) come to dominate in this regime.  For example, the first term in $\theta^\sigma$ always dominates over the second since $f\to\infty$ as $\tau\to 0$. By contrast, the relative sizes of the different terms in $\theta^s$ depend on the magnitude of  $f'/f$.  

For clarity, we define the symbol ``$\gtrsim$'' to mean
\begin{equation}
|A(\tau)|~\gtrsim~ |B(\tau)| \; \text{if}\; 
\frac{d \ln |A(\tau)|}{d\ln (-\tau)}<\frac{d\ln |B(\tau)|}{d\ln (-\tau)}\;  
\text{as}\; \tau \to 0. 
\end{equation}
Similarly, we define ``$\sim$'' to mean
\begin{equation}
A(\tau) \sim B(\tau) \; \text{if}\quad
\frac{d \ln |A(\tau)|}{d\ln (-\tau)}=\frac{d\ln |B(\tau)|}{d\ln (-\tau)}\;
\text{as}\; \tau\to 0.
\end{equation}  
For example, $1/(-\tau)^2 \gtrsim 1/(-\tau)$ and $3/(-\tau) \sim 2/(-\tau)$.

There are three qualitatively different cases to consider:  fast growth, $|f,_\sigma/f| \gtrsim 1$, slow growth, $|f,_\sigma/f| \lesssim 1$, and ``just-so'' growth, $|f,_\sigma/f| \sim 1$.  
\subsection{Fast growth:  $|f,_\sigma/f|\gtrsim1$}
\label{fastgrowth}
If $|f,_\sigma/f|\gtrsim1$, then $|f'/f|\gtrsim 1/(-\tau)$.  Let us assume that 
\begin{equation}
\label{assump}
\gamma\, \frac{u_\sigma}{(-\tau)^2},\, \delta\, \frac{u_\sigma'}{(-\tau)} \ll \theta^s \frac{u_s}{(-\tau)^2}
\end{equation}
 so that to leading order, the entropic mode evolves independently:
\begin{equation}
u_s''-\frac{1}{2}\left(\frac{f''}{f}-\frac{1}{2}\left(\frac{f'}{f}\right)^2\right)u_s=0.
\end{equation}
For the mode function $u_s$, we find the solution
\begin{equation}
\label{entlead}
u_s(\tau)=\sqrt{f(\tau)}\left( c_1(k)\int_{-1/k}^\tau \frac{d\bar{\tau}}{f(\bar{\tau})}+c_2(k) \right),
\end{equation}
where $c_1(k)$ and $c_2(k)$ are constants of integration. Choosing $c_1(k)$ and $c_2(k)$ so that $u_s$ and $u_s'$ match the Bunch-Davies  solution, $(1/\sqrt{2k})e^{-i k\tau}$, at horizon crossing, $- \tau= 1/k$, yields
\begin{eqnarray}
\label{consts1}
c_1(k) &=& \frac{e^i \left(f'\left(-1/k\right)+2 i k f\left(-1/k\right)\right)}{2 \sqrt{2k f \left(- 1/k \right)}},\\
\label{consts2}
c_2(k) &=& \frac{e^i}{\sqrt{2kf\left(-1/k\right)}}.
\end{eqnarray}
For fast-growing $f$, at late times the integral in Eq.~\eqref{entlead} is very closely approximated by $(1/k) (1/f(-1/k))$.  With Eqs.~\eqref{consts1} and \eqref{consts2}, the entropic mode function is given by 
\begin{equation}
u_s = \Sigma(k) \sqrt{f(\tau )}
\end{equation}
where 
\begin{equation}
\Sigma(k) = \frac{e^i \left(-f'\left( -1/k \right)+(2-2 i) k f\left(- 1/k \right)\right)}{2 \sqrt{2} k^{3/2} (f\left(- 1/k\right))^{3/2}}.
\end{equation}
Substituting this result into the right side of the adiabatic equation, Eq.~\eqref{modefunctionsa}, we find that  $u_\sigma$ satisfies 
\begin{equation}
u_\sigma'' + 2 \frac{ \mu ^2 - 4}{ \left(\mu ^2-2\right)^2 } \frac{1}{(-\tau)^2}u_\sigma = 4 \frac{ \mu ^2 - 6 }{\mu  \left(\mu ^2-2\right)^2 }  \frac{c\, \Sigma (k)}{(-\tau)^2}
\end{equation}
with solution
\begin{equation}
\label{adlead}
u_\sigma=c_3(k) (-\tau)^{\frac{\mu^2 - 4}{\mu^2-2}} + c_4(k)(- \tau)^{\frac{2}{\mu^2-2}}+\frac{2 \left(\mu ^2-6\right)}{\mu  \left(\mu ^2-4\right)} c \Sigma (k) ,
\end{equation}
where $c_3(k)$ and $c_4(k)$ are constants.  The first two terms vanish as $\tau\to 0$ by property~P4.  This shows that our assumption in Eq.~\eqref{assump} is justified.  

From the solutions for the mode functions in Eqs.~\eqref{entlead} and \eqref{adlead}, it is clear that both the adiabatic and the entropic spectra are proportional to $k^3|\Sigma(k)|^2$. Scale invariance is obtained if and only if
\begin{equation}
\label{diffeq}
|\Sigma(k)|^2=\xi k^{-3},
\end{equation}
for some constant $\xi$ that is independent of $k$.  Eq.~\eqref{diffeq} is a first order differential equation for the coupling function $f$, namely
\begin{equation}
\left( \frac{f'}{f} \right)^2 - \frac{4}{(-\tau)} \frac{f'}{ f}  = 8 \xi\,  f - \frac{8}{(-\tau)^2};
\end{equation}
its solution is given by
\begin{equation}
f(\tau)=\frac{\sec^2 \left[ \ln (\tau/\tau_0) \right]}{2 \xi (- \tau)^2},
\end{equation}
where $\tau_0$ is an integration constant.  This is clearly not monotonic as $\tau\to 0$ and, therefore, violates property~P1.  Hence, we conclude scale-invariance is impossible for fast-growing $f$.  

\subsection{Slow growth: $|f,_\sigma/f|\lesssim1$}
If $|f,_\sigma/f|\lesssim1$, then $|f'/f|\lesssim 1/(-\tau)$.  To leading order, both mode functions satisfy 
\begin{equation}
\label{decouple}
u'' + 2 \frac{\mu^2-4}{(\mu^2-2)^2} \frac{1}{(-\tau)^2} u = 0,
\end{equation} as is clear from Eqs.~\eqref{modefunctionsa}, \eqref{modefunctionse} and (\ref{constantterms2thsig}-\ref{constantterms2d}).  In this case, both the adiabatic and the entropic spectra are given by 
\begin{equation}
\label{adiabatic}
n_S=4-\left|\frac{\mu^2-6}{\mu^2-2}\right|,
\end{equation} 
which is blue by property~P4. Hence, we conclude scale-invariance is impossible for slow-growing $f$.

\subsection{``Just-so'' growth: $|f,_\sigma/f|\sim1$}\label{ltlimit}
If $|f,_\sigma/f|\sim 1$, $f(\sigma)=e^{-\lambda \sigma}$ for some $\lambda \in \mathbb{R}$ such that $\text{sgn}(\lambda)=\text{sgn}(\mu)$. The coupling functions $\alpha, \beta, \gamma, \delta$ are all proportional to $1/\sqrt{f}$ so the right sides of Eqs.~\eqref{modefunctionsa} and \eqref{modefunctionse} can be neglected.  The mode functions effectively decouple and the adiabatic spectral index is again given by Eq.~\eqref{adiabatic}; the entropic spectral index is  
\begin{equation}
n_S = 4 - \left| 2\frac{ \lambda\, \mu - 2}{\mu ^2 - 2} + 1\right|,
\end{equation}
which is scale-invariant when $\lambda=\mu$.  For a given $\mu>\sqrt{6}$, any $n_S<(3\mu^2-2)/(\mu^2-2)$ can be achieved by choosing $\lambda=(n_S-1)/\mu - \mu  \left(n_S-3\right)/2$.  

Note that since ``just-so'' growth implies $f(\sigma)=e^{-\lambda\sigma}$ (and property~P2 requires $h(\sigma)~=~e^{-\mu\sigma}$), the Lagrangian density is given by
\begin{equation}
\label{newlag2}
\mathcal{L}=\frac{1}{2}R-\frac{1}{2}(\partial\sigma)^2-\frac{1}{2}e^{-\lambda\sigma}(\partial\phi)^2-V_0e^{- c\phi}e^{-\mu\sigma},
\end{equation}
which is equivalent to the Lagrangian density in Eq.~\eqref{newlag3} with the identifications 
\begin{eqnarray}
\sigma &\longleftrightarrow& \psi,  \nonumber\\
\phi & \longleftrightarrow & \chi.
\end{eqnarray}

\section{Discussion}\label{discussion}

In this paper, we have presented explicit examples of ekpyrotic models, with and without shift symmetry, that have fixed-curve attractor background cosmological solutions  
generating a scale-invariant spectrum of entropic perturbations.  The existence of a negative eigenvalue associated with the linearized dynamical equations, as identified by Tolley and Wesley, can indicate a true instability requiring fine-tuning of initial conditions for some actions.  But, for actions of the type discussed here, the negative eigenvalue only indicates that the attractor solution is a curve rather than a point.  Fine-tuning of initial conditions is thus avoided.  Furthermore, as in all ekpyrotic models, this class of actions avoids the multiverse and the problem that all cosmological outcomes are possible.  Hence, the predictions are ``generic,'' the same on average for any Hubble-sized patch. 

One might be concerned if, in order to avoid fine-tuning of initial conditions, the models had to be made more complicated.  However, the opposite is the case here.  The actions impose less stringent constraints on the equation-of-state, $\epsilon$, during the contracting phase and, hence, less fine-tuning of parameters.  For such a choice of parameters, a scalar spectrum of density perturbations with the observed spectral tilt can be generated.  As in all ekpyrotic models, though, it is not possible to generate a detectable spectrum of primordial gravitational waves (the ratio of the  tensor-perturbation amplitude to the scalar-perturbation amplitude,  $r \approx 0$ ), consistent with current limits \cite{Khoury:2001wf,Boyle:2003km}. The same class of actions has also been shown to generate zero non-Gaussianity during the ekpyrotic contraction phase; a small amount of local non-Gaussianity may be generated during the bounce, but at a level well within current observational bounds on $f_{\text{NL}}$ \cite{Fertig:2013kwa,Ijjas:2014fja}.

It is notable that current cosmological observations are constraining ekpyrotic models to be in a class that is the simplest, as measured by parameters, degrees of freedom, and initial conditions.  By contrast, the same observations are pointing away from the simplest models of inflation \cite{Ijjas:2013vea}.

We thank D. Wesley and A. Tolley for useful comments.  This research was 
supported in part  by the U.S. Department of Energy under
grant number DE-FG02- 91ER40671.%
\section{Appendix}
Here we derive the four properties listed at the beginning of Sec.~\ref{sec:secGeneral} using a series of lemmas:
\begin{description}
\item[Lemma 1]  \emph{$x=0$ cannot generate a scale-invariant adiabatic spectrum without fine-tuning. }
\end{description}
If $x=0$, $z=z_0=const$.  Since $z\equiv \sigma$, this means that $\sigma \equiv z_0$ along the solution.  Defining
\begin{eqnarray}
\label{shiftsoln}
\bar{\sigma}&\equiv& \sigma- z_0,
\\
\bar{\phi}&\equiv& \sqrt{f(z_0)}\left(\phi-\frac{1}{c}\ln h(z_0)\right),
\\
\bar{c}&\equiv& \frac{c}{\sqrt{f(z_0)}},
\\
F(\bar{\sigma})& \equiv& \frac{f(\bar{\sigma}+z_0)}{f(z_0)},
\\
H(\bar{\sigma})&\equiv& \frac{h(\bar{\sigma}+z_0)}{h(z_0)},
\end{eqnarray}
 the Lagrangian density can be recast as 
\begin{eqnarray}
\label{lemma1t}
\mathcal{L}= \frac{1}{2}R-\frac{1}{2}(\partial\bar{\sigma})^2-\frac{1}{2}F(\bar{\sigma})(\partial\bar{\phi})^2-V_0H(\bar{\sigma})e^{-\bar{c}\bar{\phi}}.
\end{eqnarray}
Eq.~\eqref{lemma1t} has the shift-symmetric form of Eq.~\eqref{wtlag} with the solution $(x,z)=(0,z_0)$ at $(x,\bar{z})=(0,0)$ (with $\bar{z}\equiv\bar{\sigma}$) for which Tolley and Wesley proved that any solutions of interest are unstable. For fixed-point solutions like this (as opposed to fixed-curve solutions) instability implies the need for finely-tuned initial conditions. Thus, we conclude $|\sigma|~\to~\infty$; we are forced to \emph{fixed-curve} solutions.
\begin{description}
\item[Lemma 2]\emph{$f\to const$ cannot generate a scale-invariant adiabatic or entropic spectrum without fine-tuning.}
\end{description}
When $f=f_0=const$, the non-canonical coupling becomes canonical. Then, the background Eqs.~\eqref{w} and \eqref{x} give $(w,x)=(w_0,x_0)$ with $x_0\neq 0$ (cf. Lemma 1).  Then $h=e^{-\sqrt{6}x_0\sigma}$. In such canonically-coupled theories, the adiabatic perturbation decouples from the entropic perturbation and has a blue tilt. More precisely, the equations of motion to the Lagrangian density
\begin{equation}
\mathcal{L}=\frac{1}{2}R-\frac{1}{2}(\partial \sigma)^2-\frac{1}{2}f_0(\partial\phi)^2-V_0e^{-c\phi}e^{-d \sigma}
\end{equation}
linearized around $(w_0,x_0)=( d/\sqrt{6}, c/\sqrt{6f_0} )$ yield the same equation for both the adiabatic and the entropic mode functions 
\begin{equation}
u'' + 
\left(k^2 + \frac{2 f_0 \left(c^2+\left(d^2-4\right) f_0\right)}{
\left(c^2+\left(d^2-2\right) f_0\right)^2} \frac{1}{(-\tau)^2} 
   \right) u=0
\end{equation}
so that the spectral index is given by 
\begin{equation}
n_S= 4-\left|\frac{c^2+\left(d^2-6\right) f_0}{c^2+\left(d^2-2\right) f_0}\right|.
 \end{equation}
Avoiding fine-tuning, {\it i.e.,} negative eigenvalues in the linearized equations of motion, requires $c^2/f_0 + d^2 > 6.$  In this regime the spectral indices are blue for both the adiabatic and entropic spectra.  
\begin{description}
\item[Lemma 3] \emph{If $w\neq 0$ then $f \to (\infty$ or $const$)  at late times.}
\end{description}
For scaling solutions, the background equation for $w$, Eq.~\eqref{w}, can be recast as 
\begin{equation}
\label{f}
f,_z=A \sqrt{f} \left( \sqrt{f}-B\right),
\end{equation}
with $A\equiv\sqrt{6}\left(w^2+x^2-1\right)/x $ and $B\equiv c/w$.  If $A<0$, $x<0$ such that $z$ increases with decreasing $N$; if $A>0$, $x>0$ and $z$ decreases.  Hence, $\text{sgn}(f')=-\text{sgn}(f,_z)$.  From the right side of Eq.~\eqref{f}, we conclude that, if $B<0$, $f\to 0$ at late times and that, if $B>0$, $f\to (B^2$ or $\infty$) at late times.
With Lemma 2, we only have to consider the case $f\to \infty$.  
\begin{description}
\item[Lemma 4]\emph{ If $f\to\infty$, then $w=0$ at late times.}
\end{description}
If $f\to \infty,$ the background equation for $w$ in Eq.~\eqref{w} becomes
 \begin{equation}
w,_N = 3w\left((x^2+w^2-1)-\frac{f,_z}{f}\frac{x}{\sqrt{6}}\right).
\end{equation}
Assume $w\neq 0$.  Then $f,_z/f = \sqrt{6}(x^2+w^2-1)/x$ and, hence, $f(\sigma)=e^{-\lambda \sigma}$ with $\lambda = -\sqrt{6}(x^2+w^2-1)/x.$ In particular, $\lambda$ and $x$ must have opposite sign.  From $z,_N =\sqrt{6}x$, we see, though, that $x$ and $z$ must have different sign (since $N$ decreases as the system evolves, $z$ will, for example, \emph{decrease} when $z,_N>0$).  Therefore, $\lambda$ and $z$ must have the same sign. But then $f=e^{-\lambda z}\to 0$ at late times, which contradicts our assumption, $f\to \infty$.

To this point, we have shown that any ekpyrotic, scaling attractor that generates a scale-invariant spectrum of either adiabatic or entropic perturbations lies at $w=0$ with $f\to \infty$ and $x\neq 0$.  Defining $\mu\equiv\sqrt{6}x$, Eq.~\eqref{x} implies $h,_z/h=-\mu$, from which we conclude, given that $z\equiv \sigma$, $h(\sigma)\propto e^{-\mu\sigma}$ (property~P2). 
\begin{description}
\item[Lemma 5]\emph{The scaling solution must correspond to the limit given in property~P3.}
\end{description}
With $h(\sigma)=e^{-\mu\sigma}$, the only scaling solutions of Eqs.~\eqref{w} and \eqref{x} are $(w,x)=\left\{(0,\pm1), (0,\mu/\sqrt{6})\right\}$ if $f,_\sigma/f \neq const$.  For the first two solutions, $\epsilon = 3$. Since ekpyrosis  corresponds to $\epsilon>3$, we only consider the third solution (property~P3).

Note, if $f,_\sigma/f =const\equiv-\lambda$, there is another scaling solution at 
\begin{equation}
\label{special}
(w,x)=\left(\pm\frac{\sqrt{\mu  (\lambda +\mu )-6}}{\left| \lambda +\mu \right| },\frac{\sqrt{6}}{\lambda+\mu}\right).
\end{equation} 
At late times, this solution is equivalent to one in Ref.~\cite{DiMarco:2002eb} that was shown never to be an attractor.
\begin{description}
\item[Lemma 6] \emph{The coupling $f(\sigma)$ must grow monotonically.}
\end{description}
Since the solution in property~P3 is a fixed-curve, it can be parameterized by one variable, $z$. For any finite $|z|$, $f$ will be finite so that even if the system lies at $(w,x)=(0,\mu/\sqrt{6})$, the kinetic energy of the ekpyrotic field evolves as $w,_N= - 3\,c\,( \mu^2/6 - 1)\sqrt{f}$ (cf. Eq.~\eqref{w}).  Thus, in any interval over which $f$ shrinks, $|w,_N|$ grows.  For this reason, we only consider solutions for which $f$ grows monotonically (property~P2).
Linearizing Eqs.~\eqref{w} and \eqref{x} about the background in property~P3 yields 
\begin{eqnarray}
&\delta w ,_N&=\left(3\left(\frac{\mu^2}{6}-1\right)-\frac{\mu}{2}\frac{ f,_z}{f}\right)\delta x,\\
&\delta x,_N&=3\left(\frac{\mu^2}{6}-1\right)\delta x,
\end{eqnarray}
where
$\delta w\equiv w$ and $\delta x\equiv x-\mu/\sqrt{6}$.  The eigenvalues of this system are $\left \{ (\mu^2-6 )/2,  \left( \mu^2 - 6 - \mu f,_z/f \right)/2 \right\}$. Since $\mu f,_z/f<0$, this reduced $2\times 2$ system is stable, {\it i.e.}, has positive eigenvalues, if $|\mu|>\sqrt{6}$ (property~P4).  

\bibliographystyle{apsrev4-1}
\bibliography{bibliography}
\end{document}